\documentclass[aps,prl,showpacs,twocolumn]{revtex4}
\usepackage{graphicx}
\usepackage{amsmath}
\newcommand{\beq}{\begin{equation}}
\newcommand{\eeq}{\end{equation}}
\newcommand{\bea}{\begin{eqnarray}}
\newcommand{\eea}{\end{eqnarray}}
\newcommand{\pdag}{{\phantom{\dagger}}}

\begin{document}
\title{ Shot noise free conductance reduction in quantum wires }
\author{M.\ Kindermann and P.\ W.\ Brouwer}
\affiliation{ Laboratory of Atomic and Solid State Physics, Cornell University, Ithaca, New York 14853-2501  }

\date{June 2005}
\begin{abstract}
We show that a shot noise free current at conductance below $2 e^2/h$ is possible in short interacting
 quantum wires   without
spin-polarization. Our calculation is done for two exactly solvable
limits of the ``Coulomb Tonks gas'', a one-dimensional
gas of impenetrable electrons that can be realized in ultra-thin 
quantum wires. In both cases we find that charge transport through 
such a wire is noiseless at zero temperature while the conductance 
is reduced to $e^2/h$.  
 \end{abstract}
\maketitle

The current through short electrical conductors is generically noisy
even in the absence of thermal fluctuations. This nonequilibrium noise
is usually referred to as shot noise and originates from
backscattering of electrons in the conductor \cite{Bla00}.  The model
of noninteracting electrons predicts that shot noise can only be
avoided in fully ballistic wires without scatterers for electrons. The
conductance of such wires is quantized in multiples of the
conductance quantum $2 e^2/h$ \cite{vWees88,Wha88}. This implies that the 
electrical current through a noninteracting wire with conductance 
below $2 e^2/h$ is bound to be noisy unless one polarizes the spins 
of electrons. The suppression of shot noise for ballistic wires with
conductance equal to a multiple of $2 e^2/h$ has been verified experimentally
for quantum point contacts \cite{Rez95}. Recently, however,
experiments on similar conductors have shown that shot noise can 
be suppressed as well at conductance $\simeq 0.7
\times 2 e^2/h$, {\em i.e.}, below the ballistic conductance minimum
$2 e^2/h$ \cite{noise}. 


The measured suppression cannot be understood  within the model of noninteracting
electrons unless one assumes the breaking of a symmetry, e.g. by spin-polarization \cite{noise}. Hence, one
may ask whether it is possible to have
noise-free conduction at a conductance below $2 e^2/h$ as a result of electron-electron interactions, but  without spin-polarization. In this letter we show that the answer
is positive. For a specific theoretical model, the strongly interacting  ``Coulomb
Tonks gas'' of impenetrable electrons \cite{Fog05,Fiete05},
we find a noiseless charge current in a wire with conductance  $0.5 \times 2 e^2/h$. It  is
accompanied by pronounced spin current fluctuations, which precludes 
a spin-polarization of electrons as the mechanism of 
conductance reduction. Our calculation is done for a finite-length 
wire in two exactly solvable limits: a high applied bias, and a
fast spin-relaxation rate. The Coulomb Tonks gas can be realized
in ultra-thin wires, such as carbon nanotubes \cite{Fog05}, for 
which there exists some experimental evidence of an anomalous 
conductance reduction \cite{CN}. 

Theoretically it has been shown by Matveev that the low-bias conductance of a quantum wire is reduced to
$e^2/h$ at low electron
density, when interactions induce the formation of a Wigner crystal,
  if the temperature is
larger than the typical energy $J$ of spin excitations \cite{Mat04}. This  is one of the scenarios that may 
explain the anomalous conductance reduction observed experimentally 
in quantum point contacts \cite{07}. The Coulomb Tonks gas studied
here is closely related to Matveev's model at $J=0$, since
impenetrable electrons have no exchange interaction.

We model a wire in the Coulomb Tonks gas regime by a Hubbard chain 
with infinite on-site repulsion, $U\to \infty$. For an infinite-length
Hubbard chain, this model can be formulated in terms of spinless holes
and a static spin background. As the only charge carriers are
spinless fermions, one expects that such a chain has a reduced 
conductance of $e^2/h$ without shot noise. Nanoscale
conductors studied in experiments, however, have a finite length and
they are connected to bulk leads with well-screened interactions. As 
the example of Luttinger Liquids shows, these contacts can nullify 
interaction effects on the conductance \cite{LL} as well as the 
shot noise \cite{LLnoise}   in infinite wires. Once
coupled to noninteracting reservoirs, the spin background in the
infinite-$U$ Hubbard chain acquires nontrivial dynamics because of
spin-exchange processes of the chain with the reservoirs, making the
model hard to solve. Below, we will identify two limits in 
which this spin dynamics can be controlled and the transport
properties of the system can be evaluated exactly.


\begin{figure}
\includegraphics[width=8.5cm]{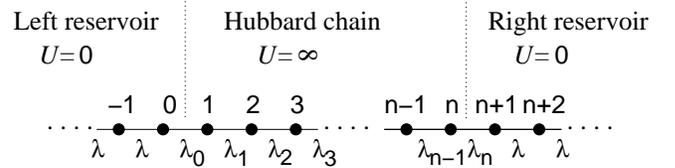}
\caption{  Hubbard chain with sites $j$ coupled to electron reservoirs $L$ and $R$. The hopping amplitude $\lambda_j$ varies along the chain and there is a repulsive on-site  interaction $U$.  }  \label{fig1} 
\end{figure}

The system we consider is shown in Fig.\ \ref{fig1}. The sites of
the infinite-$U$ Hubbard chain are labeled $j=1,\ldots,n$. The Hubbard
chain is coupled to two
noninteracting leads, with sites $j \leq 0$ and $j > n$.
We introduce operators $\psi_{j \sigma}$ that annihilate an electron 
with spin $\sigma$ on site $j$ in the leads and
$c_{j}$ that annihilate a hole at position $j$ in the Hubbard
chain. The spin index $\sigma$ takes the values
$1,\ldots,2S+1$. We also introduce operators $S_{{\rm L}\sigma
}^{\dagger}$ and $S_{{\rm R}\sigma }^{\dagger}$ that add a spin $\sigma$ 
to the left and right of the spin configuration of the
Hubbard chain, respectively. Our model Hamiltonian then reads
\bea \label{H}
  H &=& H_{\rm lead} + H_{\rm chain} +  H',
\eea
with
\bea
  H_{\rm lead} &=& \lambda \sum_{\sigma}
  \left[\sum_{j < 0}
  \psi^{\dagger}_{j \sigma}
  \psi_{j+1 \sigma} 
  + \sum_{j > n} \psi^{\dagger}_{j \sigma}
  \psi_{j+1 \sigma} \right] + \mbox{h.c.} \nonumber \\
  H_{\rm chain} &=&
  \sum_{j=1}^{n-1} \lambda_j c^{\dagger}_{j} c_{j+1}
  + \mbox{h.c.}
  \nonumber \\
  H' &=&
  \sum_{\sigma} \left( \lambda_0 \psi_{0 \sigma}^{\dagger} c_1^{\dagger}
  S^{\pdag}_{{\rm L}\sigma } + \lambda_n \psi_{n+1 \sigma}^{\dagger} c_n^{\dagger}
  S^{\pdag}_{{\rm R}\sigma } \right) + \mbox{h.c.}, \nonumber
\eea
where we allowed for spatial variations of the hopping amplitude
$\lambda_j$ of the Hubbard chain. 

Moments   of the number $\Delta N_{\sigma}=N_\sigma(\tau)-N_\sigma(0)$ of spin $\sigma$ electrons that are transported through the chain during a time $\tau$ can be obtained from the generating function  
\beq  
 {\cal Z}(\{\xi_{\sigma}\})= {\rm Tr} \, e^{-i \xi N }\, e^{-iH\tau}\, e^{i  \xi  N} \,\rho\, e^{iH\tau}.
 \eeq
Here
 $\rho$ is the  initial density matrix of the conductor  and $\xi N$ abbreviates $  \sum_\sigma  \xi_\sigma N_\sigma$ with  $N_\sigma=  \sum_{j\leq 0}\psi^{\dag}_{j \sigma}\psi^{\pdag}_{j \sigma}$.
For large $\tau$,  ${\cal Z}$ generates zero-frequency
(spin)
 current correlators  upon differentiation.  We rewrite ${\cal Z}$ as
\beq \label{Z}
 {\cal Z}(\{\xi_{\sigma}\})= {\rm Tr} \, e^{-iH_{\xi}\tau}\, \rho\,  e^{iH\tau}, 
\eeq
where $H_{\xi} = \exp({-i \xi N })H \exp({i \xi N})$. $H_{\xi}$ is obtained from $H$ by the substitution
$\psi_{j\sigma} \to   e^{i \xi_\sigma }\psi_{j\sigma}$ for $j\leq0$. Since the Hamiltonian (\ref{H}) is
quadratic in the fermions, the lead fermions may be integrated out,
so that the generating function ${\cal Z}$ takes the form
\begin{widetext}
\bea
  {\cal Z}(\{\xi_{\sigma}\}) &=&
  \left\langle T_{\rm c} 
    \exp\left[- i  \lambda_0^2 \int_{\rm c} dt_1 dt_2 \sum_{\sigma}
    S^{\dagger}_{{\rm L} \sigma}(t_1)
    c_1(t_1) G_{{\rm L} \sigma \xi}(t_1,t_2) c^{\dagger}_1(t_2) S^{\pdag}_{{\rm L} \sigma}(t_2)
  \right. \right. \nonumber \\ && 
  \hphantom{\left\langle T_{\rm c} 
    \exp\left[ \vphantom{\int} \right. \right.}
  \left. \left. \mbox{}
  \!\!\!\!  - i  \lambda_n^2\int_{\rm c} dt_1 dt_2 \sum_{\sigma}
     S^{\dagger}_{{\rm R} \sigma}(t_1)
    c_n(t_1) G_{{\rm R} \sigma}(t_1,t_2) c^{\dagger}_n(t_2) S^{\pdag}_{{\rm R} \sigma}(t_2)\right]
  \right\rangle_{\rm chain},
  \label{Zeff}
\eea
\end{widetext}
where c denotes the Keldysh contour, $G_{{\rm L}}$ and $G_{{\rm R}}$ are
Keldysh Green functions for the sites $0$ and $n+1$, respectively,
and the averaging brackets $\langle \ldots \rangle_{\rm chain}$ represent
an average with respect to the Hamiltonian $H_{\rm chain}$
of the uncoupled Hubbard chain. Using matrix notation
in Keldysh space, one has $G_{{\rm L}\sigma \xi} = e^{i \xi_{\sigma} \tau_3/2} G_{{\rm L}\sigma}
e^{-i \xi_{\sigma} \tau_3/2}$, where $\tau_3$ is the third Pauli matrix in Keldysh space.

The first limit in which exact results can be obtained is that 
of reservoirs with a large bandwidth $\lambda\to \infty$ that are 
completely filled with electrons on the left and completely empty 
on the right side of the chain. 
In that case, the reservoir Green 
functions $G_{{\rm L}}$ and $G_{{\rm R}}$  take a particularly simple form,
\bea \label{GLs}
G_{{\rm L}}(t,t')&=&\frac{2i}{\lambda}\delta_\lambda(t-t')  \left(
\begin{array}{cc}  \theta(t'-t)&   1 \\ 0& \theta(t-t')  \end{array} \right) ,   \\
 \label{GRs} 
G_{{\rm R}}(t,t')&=&-\frac{2i}{\lambda}\delta_\lambda(t-t')  \left(
\begin{array}{cc}  \theta(t-t')&  0\\  1 & \theta(t'-t)  \end{array} \right) ,
 \eea 
where $\delta_\lambda$ is an approximation to a Dirac delta-function 
with width $1/\lambda$. 
 In order to be able to take the limit $\lambda\to \infty$ while
 having a finite transport current we assume that the bandwidth of the
 chain is adiabatically reduced.  We assume that $\lambda_j = \lambda$
 for $j$ close to the boundaries of the Hubbard chain at $j= 1$ and
 $j=n$, 
and that $\lambda_j$ is adiabatically
reduced to a value $\lambda_j=\lambda_{\rm r} \ll \lambda$ in the center of the chain, 
$j \simeq n/2$.


Below, we will argue that, in this  limit,
one may make the following substitutions in the exponent of
Eq.\ (\ref{Zeff}),
\begin{eqnarray}
 \sum_{\sigma}  S_{{\rm L} \sigma}^{\dagger}(t_1) 
G_{{\rm L}\sigma\xi}(t_1,t_2)S^{\pdag}_{{\rm L} \sigma}(t_2) & \to & \sum_\sigma G_{{\rm L}\sigma\xi}(t_1,t_2), \nonumber
  \\
  \sum_{\sigma} S_{{\rm R} \sigma}^{\dagger}(t_1) 
  G_{{\rm R}\sigma}(t_1,t_2)  S^{\pdag}_{{\rm R} \sigma}(t_2)& \to & G_{{\rm R}\sigma}(t_1,t_2).
  \label{repl}
\end{eqnarray}
After these replacements,
the generating function ${\cal Z}$ is readily calculated, and we
find
\beq \label{Zr}
\ln {\cal Z}= \frac{\tau}{h} \int_{-2 \lambda_{\rm r}}^{2 \lambda_{\rm r}}
  {d\omega 
  \ln\left[1+ \frac{T(\omega)}{2 S+1} \sum_\sigma{(e^{i\xi_\sigma}}-1)
  \right]},
\eeq
where, for $\lambda_{n}=\lambda$,
\beq \label{T}
T(\omega)= \frac{(2S+1) |\lambda_0|^2(4\lambda^2-\omega^2)}{(\lambda^2+(2S+1)|\lambda_0|^2)^2-(2S+1)|\lambda_0|^2 \omega^2}.
\eeq
(We set $\lambda_n=\lambda$ to simplify the expressions
reported here; it is not essential for the calculation.)
The function $T(\omega)$ 
  reaches its maximum $T^{\rm max}(\omega)=1$ for 
$\lambda_0=\lambda/\sqrt{2S+1}$ \cite{arti}.    While the maximum current in the corresponding wire without
  interactions  would be
  $I^{\rm max}_{\rm n}=-4 (2S+1)\lambda_{\rm r} \,e/h$, Eq.\ (\ref{Zr}) implies for the interacting case
\beq
 I^{\rm max}= \frac{1}{2S+1} I^{\rm max}_{\rm n}.
\eeq
Hence, for electrons with spin $S=1/2$ interactions  reduce the maximal charge current by a factor of $2$. This signals a reduction of the conductance of the wire from $2 e^2/h$ to $e^2/h$, as in the infinitely long system. 
 Every spin component carries  an equal part of that total charge current independently of the chosen spin quantization direction: the electrons in the wire are not spin-polarized. This absence of spin-polarization in the wire is also reflected in fluctuations of the spin currents through the wire.   At maximal transmission  we find for their variance
 \beq
 \mbox{var}\, \Delta N_\sigma = \frac{4\lambda_{\rm r}\tau}{h} 
  \frac{2 S}{(2 S+1)^2}.
 \eeq 
 In contrast, the charge current  is noiseless in this case: 
\beq
 {\rm var}\,\left(\sum_\sigma \Delta N_\sigma\right) =0.
 \eeq 
Our model therefore indeed exemplifies a mechanism of  conductance reduction through electron-electron interactions that does not  introduce any shot noise. Instead of blocking the transport of electrons with certain spin directions by spin-polarizing the wire, this mechanism reduces the conductance by   limiting the density of conduction electrons to the spinless case while allowing them to have arbitrary spin. 

We now discuss the justification of the replacements (\ref{repl}). 
This is done by
expanding the exponential of Eq.\ (\ref{Zeff}) in powers of $\lambda_0$
and $\lambda_n$.  For simplicity we specialize to the case
$\lambda_0 = \lambda_n/\sqrt{2 S+1}$. Since the spin operators
$S^{\pdag}_{{\rm L} \sigma}$ and $S_{{\rm L} \sigma}^{\dagger}$ have no dynamics in the
absence of coupling to the reservoirs, the time dependence of the
spin operators is not important; only their order in the
contour-ordered expression (\ref{Zeff}) matters.
To second order in $\lambda_n$, the causal 
structure of the reservoir Green function
$G_{\rm L}$ is such that $S^{\pdag}_{{\rm L}\sigma}(t_2)$ always appears to the 
left of $S_{{\rm L}\sigma}^{\dagger}(t_1)$. Since $S^{\pdag}_{{\rm L}\sigma} 
S_{{\rm L}\sigma}^{\dagger} = 1$, the first substitution rule
(\ref{repl}) follows.
Similarly, the causal structure of the reservoir Green function
$G_{\rm R}$ is such that $S^{\pdag}_{{\rm R}\sigma}(t_2)$ always appears to the 
right of $S_{{\rm R}\sigma}^{\dagger}(t_1)$. The operator product 
$S^{\dagger}_{{\rm R} \sigma} S^{\pdag}_{{\rm R}\sigma} = 1$ if the right-most electron
of the Hubbard chain has spin $\sigma$, and $S^{\dagger}_{{\rm R} \sigma} 
S^{\pdag}_{{\rm R}\sigma} = 0$ otherwise. Hence $\sum_{\sigma} S^{\dagger}_{{\rm R}
  \sigma} S^{\pdag}_{{\rm R}\sigma} = 1$, and the second substitution rule of
Eq.\ (\ref{repl}) follows.

We next prove the substitution rules (\ref{repl}) for higher orders
in  $\lambda_n$. A key ingredient in this will be  the locality of the reservoir Green functions which groups spin operators  into ordered pairs that act  almost simultaneously. This essentially decouples different spin operator  pairs and the rules (\ref{repl}) follow like at first order in $\lambda_n$. To show  this we decompose the hole
annihilation operators $c_j=c_{j,e}+c_{j,t}$ into a contribution
$c_{j,e}$ from evanescent states of the chain at energies 
$\epsilon\geq 2\lambda_{\rm r}$ and a contribution $c_{j,t}$ from
transmitting states with $\epsilon<2\lambda_{\rm r}$. In any term of an
expansion of ${\cal Z}$ in powers of  $\lambda_n$ we
first contract all hole operators $c_{j,e}$ into pairs using Wick's
theorem. 
This produces sequences of the form
\begin{eqnarray}
  D_{{\rm L};l} &=& 
  \int_{\rm c} dt_2 \ldots dt_{l-1}\,
  c^\pdag_{1,t}(t_1)
  S^\dag_{\rm L}(t_1) G_{{\rm L}\xi}(t_1,t_{2})
  \nonumber \\ && \mbox{} \times
  S_{\rm L}^\pdag(t_{2}) g_e(t_{2},t_{3})
  \ldots S_{\rm L}^\pdag(t_{l}) c^\dag_{1,t}(t_{l}),
  \label{eq:sequence}
\end{eqnarray}
consisting of spin operators and two transmitting hole operators $c_{1,t}$,
connected by Green functions for lead and chain fermions. 
All evanescent hole states on the left side of the chain are
unoccupied because of their proximity to the left reservoir, so that
the evanescent hole Green function $g_e$ has the same causal structure
as $G_{{\rm R}}$, {\em cf.}\ Eq.\ (\ref{GRs}). 
This assures that after time ordering all spin operators in such
sequences are grouped in pairs $S^\dag_{\rm L}(t_i) G_{{\rm L}\xi}(t_i,t_{i+1})
S_{\rm L}^\pdag(t_{i+1}) $ originating from the same reservoir Green function
with no other spin operator $S^\pdag_{\rm L}$ or $S^\dag_{\rm L}$ acting in between. The
substitution rules (\ref{repl}) follow then for such sequences as
before for a single pair. 
The same arguments hold for the right side of the chain. 
It remains to contract the operators $c_{j,t}$ at the beginning
and at the end of sequences $D_{{\rm L},{\rm R},l}$ of the form (\ref{eq:sequence}). 
Here, nothing prevents spin operators of one sequence $D_{{\rm L},{\rm R};k}$
to occur in time between operators $S^\dag_{\rm L}(t_i) G_{{\rm L}\xi}(t_i,t_{i+1}) 
S_{\rm L}^\pdag(t_{i+1})$ in another sequence $D_{{\rm L},{\rm R};l}$. Such events cause 
corrections to the generating 
function calculated with the substitution rules (\ref{repl}). However,
because of the locality of the lead Green functions $G$, such corrections 
occur only in time intervals of length $1/\lambda$. This is much
shorter than the range of the integrals over the times $t_1$ and $t_l$ of sequences $D_{{\rm L},{\rm R};l}$, which is set by the time scale $1/\lambda_{\rm r}$ on which the Green
functions of transmitting hole states vary. Therefore, corrections
are of order $\lambda_{\rm r}/\lambda$, and can be
neglected in our limit $\lambda_{\rm r} /\lambda \to 0$. 

To make this argument rigorous one needs to estimate the corrections
to all terms in an expansion of the cumulant generating function
$\ln{\cal Z}$. Only ``connected'' diagrams contribute to 
 $\ln{\cal Z}$. Spin operators in its expansion can be connected 
either by fermion Green functions or by deviations from the operator ordering 
underlying the substitution rules (\ref{repl}). As we argued above, 
the corrections to diagrams where all spin operators are
connected by Green functions are of order $\lambda_{\rm r}/\lambda$. 
Let us now estimate the correction arising 
from diagrams that are connected because of deviations from the  ordering
underlying the rules (\ref{repl}). Hereto,
note that a product of two sequences $j$ and $k$ of spin operators that are connected cyclically by fermion Green functions only within themselves still contributes
to $\ln{\cal Z}$ if a spin operator of one of the sequences acts in
between an operator pair $S^\dag_{\rm L}(t_i) G_{{\rm L}\xi}(t_i,t_{i+1})
S_{\rm L}^\pdag(t_{i+1})$ of the other sequence. When uncorrelated the two 
subsequences are of order $\tau (\lambda_n/\lambda)^{2n_j} \lambda_{\rm r}$ 
and $\tau (\lambda_n/\lambda)^{2n_k} \lambda_{\rm r}$, respectively, where 
$n_m$ is the number of reservoir Green functions $G$ contained in 
sequence $m$. Correlations between the subsequences occur for the time 
$1/\lambda$ over which spin operators interfere and the corresponding 
correction is therefore of order 
$\tau(\lambda_n/\lambda)^{2n_j+2n_k}(\lambda_{\rm r}/\lambda) \lambda_{\rm r}$.
This is of the same order as the correction to a 
sequence connected entirely by Green functions at the same order in $\lambda_n$. This correspondence holds for diagrams with an arbitrary 
number of connections by spin operators. We now take into account that
every one of the discussed sequences contains many pairs of spin 
operators that can interfere with other operators.  We estimate that 
at order $(\lambda_{\rm r}/\lambda)^k$ 
the correction $\Delta_{k;l}$ to a term of order $2l$ in $\lambda_n$ in 
the expansion of $\ln{\cal Z}$ is of relative order 
\beq
 \Delta_{k;l} \sim \left(\begin{array}{cc}  2l \\  k  \end{array} \right) \left(\frac{\lambda_n}{\lambda}\right)^{2l}   \left(\frac{\lambda_{\rm r}}{\lambda}\right)^k.
 \eeq
Summing up these contributions one finds that the total correction to
$\ln {\cal Z}$ is  of relative order $\epsilon\ll 1$ if $\lambda_n/\lambda <
\exp(\lambda_{\rm r} \ln \epsilon/\lambda \epsilon)$. In the limit
$\lambda_{\rm r}/\lambda \to 0$ the substitution rules Eq.\ (\ref{repl}) are
therefore justified for arbitrarily good transmission $T < 1$, where
$T \to 1$ in the limit $\lambda_n \to \lambda$.    
 
The second limit in which exact results can be obtained is in 
the presence of additional spin relaxation processes in the Hubbard
chain. To the Hamiltonian (\ref{H}) we add a random time-dependent
magnetic field that causes spin memory to
be lost at a rate $\gamma_S$. In the limit $\gamma_S \gg |\lambda|$
our model is again exactly solvable, now both in and out of 
equilibrium and without assumptions about $\lambda_j$. To demonstrate 
this we consider  a wire with a constant  $\lambda_j =\lambda$. 
Every operator $S^{\pdag}_{{\rm R}\sigma}(t)$ or $S^{\pdag}_{{\rm L}\sigma}(t)$ in 
an expansion of ${\cal Z}$ is nonvanishing only if $\sigma$ equals
the state of the last or first spin of the Hubbard chain at time $t$. In the limit
$\gamma_S \gg |\lambda|$ any memory of the state of the spin put into 
the chain by previous actions of the operators
$S^{\dag}_{{\rm R}\sigma}$ and $S^{\dag}_{{\rm L}\sigma}$
is immediately erased by relaxation processes. After averaging  over the random magnetic field, every term from the exponent of Eq.\ (\ref{Zeff}) contributes therefore only for one random spin state $\sigma$. This allows us again 
to represent  ${\cal Z}$ by an expression of the form of Eq.\
(\ref{Zeff}) without spin operators. To obtain the 
generating function of charge currents we set $\xi_\sigma=e\xi$ and evaluate Eq.\ (\ref{Zeff}) with the substitutions 
\begin{eqnarray}
 \sum_{\sigma} S_{\mu \sigma}^{\dagger}(t_1) 
G_{\mu\sigma}(t_1,t_2) S_{\mu \sigma}(t_2) & \to & G_{\mu\sigma}(t_1,t_2),  
  \label{repldecoh}
\end{eqnarray}
$\mu \in \{{\rm L},R\}$, to find that
 \bea \label{Zdecoh}
\ln {\cal Z}&=& \frac{\tau}{h} \int_{-2\lambda}^{2\lambda}{d\omega\,  \ln\left\{1+T(\omega)\right.} \nonumber \\
&&\!\!\!\!\!\!\!\!\!\!\!\!\!\!\!\!\!\!\!\!\!\!\!\!\!\!\!\!\!\!\!\! \left. \mbox{}\;\left[f_L(1-f_R)\left(e^{i e\xi}-1\right) +f_R(1-f_L)\left(e^{-ie\xi}-1\right)\right]\right\}.
\eea
Here, $T(\omega)$ is given by  Eq.\ (\ref{T}) evaluated at $S=0$ and
 $f_\mu$ are the occupation numbers of reservoir states.   $T(\omega)$
 now reaches its  maximum $T^{\rm max}(\omega)=1$ for
 $\lambda_0=\lambda_n=\lambda$.  Charge transport through the chain
 becomes indistinguishable from  that through a  noninteracting
 single-channel conductor for spinless fermions. We again find
 noiseless  charge current  at zero temperature while the conductance
 is reduced to $e^2/h$, a factor $2S+1$ smaller than its noninteracting value. This shows that the discussed mechanism of noiseless conductance reduction works also in equilibrium and for a not fully occupied electronic band. In fact, the condition for the spin relaxation rate $\gamma_S$ may be relaxed for a wire with an almost empty conduction band. For a long wire, $n\gg \lambda^2/\Delta\lambda\gamma_S$,  Eq.\ (\ref{Zdecoh}) can be shown to hold also in the experimentally more relevant situation that the spin decoherence rate in the wire exceeds the range  $\Delta \lambda$ of energies for which electronic states are appreciably occupied, $\gamma_S \gg \Delta\lambda$ \cite{unpub}.    

In conclusion, by studying two exactly solvable limits of a Coulomb
Tonks gas coupled to bulk leads, we have shown that strong electron-electron interactions
can lead to a noiseless conductance reduction without spin-polarization.
One may speculate that the same mechanism could be operative also in
other parameter regimes. As such it  appears to be a plausible
alternative interpretation of the measurement of a suppression of shot
noise at the 0.7-plateau of quantum point contacts reported in  Refs.\
\cite{noise}.  Further evidence for the studied mechanism could be
found in similar measurements on carbon nanotubes, to which our model
applies more directly \cite{Fog05}. Such measurements could moreover
put an important constraint on theories of the  anomalous
conductance reduction observed in some quantum wires \cite{07,CN}.


This work was supported by the NSF
under grant no.\ DMR 0334499 and by the Packard Foundation.


\begin{thebibliography}{99}
\bibitem{arti}  The discontinuous coupling
$\lambda^{\rm max}_L=\lambda/\sqrt{2S+1}$ at maximal transmission is an
artifact of our model. It would be smoothed out if interactions were decreased adiabatically towards the
reservoirs.
\end{thebibliography}
\end{document}